\documentclass[conference]{IEEEtran}
\IEEEoverridecommandlockouts
\usepackage{cite}
\usepackage{amsmath,amssymb,amsfonts}
\usepackage{algorithmic}
\usepackage{algorithm}
\usepackage{listings}
\usepackage{graphicx}
\usepackage{textcomp}
\usepackage{xcolor}
\usepackage{amsmath}
\usepackage{bbding}
\usepackage{multirow}
\usepackage{array}
\usepackage{alltt}
\usepackage{listings}
\usepackage{diagbox}

\usepackage{ulem}

\usepackage{framed}
\usepackage{graphicx}
\usepackage{subcaption}

\usepackage{amsmath}

\usepackage{tabularx}

\usepackage{caption}
\usepackage{titlesec}
\titlespacing*{\subsection}{0pt}{5pt}{5pt}

\usepackage{url} 
\usepackage{booktabs}
\def\BibTeX{{\rm B\kern-.05em{\sc i\kern-.025em b}\kern-.08em
    T\kern-.1667em\lower.7ex\hbox{E}\kern-.125emX}}

\begin{document}

\title{Gensor: A Graph-based Construction Tensor Compilation Method for Deep Learning}

\author{
\IEEEauthorblockN{Hangda Liu\IEEEauthorrefmark{2}\IEEEauthorrefmark{3}, 
                  Boyu Diao\IEEEauthorrefmark{2}\IEEEauthorrefmark{3}\IEEEauthorrefmark{1}, 
                  Yu Yang\IEEEauthorrefmark{2}\IEEEauthorrefmark{3}, 
                  Wenxin Chen\IEEEauthorrefmark{2}\IEEEauthorrefmark{3}, 
                  Xiaohui Peng\IEEEauthorrefmark{2}\IEEEauthorrefmark{3},
                  Yongjun Xu\IEEEauthorrefmark{2}\IEEEauthorrefmark{3}}
\IEEEauthorblockA{\IEEEauthorrefmark{1} Corresponding Author: diaoboyu2012@ict.ac.cn}
\IEEEauthorblockA{\IEEEauthorrefmark{2} Institute of Computing Technology, Chinese Academy of Sciences, Beijing, China}
\IEEEauthorblockA{\IEEEauthorrefmark{3} University of Chinese Academy of Sciences, Beijing, China}
\IEEEauthorblockA{\{liuhangda21s, diaoboyu2012, pengxiaohui, xyj\}@ict.ac.cn, \{yangyu211, chenwenxin22\}@mails.ucas.ac.cn}
}

\maketitle

\begin{abstract}
High-performance deep learning depends on efficient tensor programs. In recent years, automatic tensor program optimization, also known as tensor compilation, has emerged as the primary approach to generating efficient tensor programs. However, how to generate kernels with higher performance in a shorter time is still the key challenge. In this paper, we present Gensor, a graph-based construction tensor compilation method for deep learning, to further improve the performance of construction tensor compilation. Unlike existing tree-based methods, Gensor abstracts construction space into a graph structure. Gensor then explores the construction space with Markov analysis. Gensor takes tensor programs as states and models scheduling primitives as transition actions between these states. Therefore, the process of tensor program construction optimization is abstracted as a graph traversal process. This approach expands the optimization space, improving operator performance while ensuring rapid optimization. Extensive experiments with typical operators demonstrate that Gensor significantly outperforms the state-of-the-art methods on GPUs for both cloud servers and edge devices. As a result, Gensor can generate operator kernels in seconds, with performance increasing by 18\% on average, reaching a maximum of 30\%. It also achieves high speedup for end-to-end models like ResNet-50 and GPT-2, with an average acceleration of 20\%.
\end{abstract}


\section{Introduction}
Deep Neural Networks (DNNs) are extensively deployed in various fields such as scientific research\cite{xu2021artificial}\cite{ren2024dmss}, autonomous driving\cite{cordts2016cityscapes}, and many other AI tasks\cite{huang2024etag}\cite{liucontinual}. However, as DNN architectures are becoming more complex, the time required for inference increases significantly. Thus, many optimization techniques are proposed to accelerate the inference process, ensuring that these networks run efficiently\cite{lin2023tiny}.

The inference process of DNNs can be represented as the execution of multiple operator kernels implemented by tensor programs. Therefore, optimizing these tensor programs to improve their ability to process tensors in parallel is vital for enhancing DNN performance, which has been extensively studied\cite{li2020deep}. Generally, tensor program optimization includes manual and automatic methods. Manual optimization methods, such as CUDA Deep Neural Network Library (cuDNN)\cite{NVIDIA2023cuDNN} and Basic Linear Algebra Subprograms (cuBLAS)\cite{NVIDIA2023cuBLAS}, manually refine the code structure to accelerate tensor programs. However, the primary drawbacks of manual optimization are the requirement for expert knowledge, accompanied by a significant time investment required for development. These make it difficult to agilely support new operators. 

By contrast, automatically optimizing tensor programs, known as tensor compilation, is proposed\cite{chen2018learning}. This approach reduces manual development costs, even discovering optimizing combinations that human experts might overlook. These methods abstract the tensor program into an intermediate representation, which is called tensor IR\cite{feng2023tensorir}. Then, they tune the optimizing combinations, which consist of techniques like loop unrolling and caching, to find tensor programs with higher parallelism, leading to superior performance.\cite{xia2023optimizing}.

The tensor compilation is generally classified into two categories: searching methods and constructive methods\cite{bi2022balto}. Searching methods utilize algorithms to explore an expansive optimizing combination space with a learning policy. These methods, like Ansor\cite{zheng2020ansor}, excel at unearthing solutions by iteratively combining, testing, and comparing different optimizing combinations. However, the main drawback of the searching methods is the substantial time and memory required for the exhaustive search\cite{shao2022tensor}. This drawback impedes the real-time compilation optimization of dynamic neural networks, where the network structure or input channels change dynamically\cite{lin2023tiny}. In contrast, the constructive methods directly apply predetermined rules to construct efficient tensor programs with an analysis policy. These methods, like Roller\cite{zhu2022roller}, are more time-efficient as they do not require extensive searching or actual testing on the target platform. However, the algorithms of existing construction tensor compilation methods are tree-based. As a result, these methods often become trapped in a local optimum and fail to identify the optimal solution.


Therefore, it is essential to improve these tree-based methods, developing a way that not only rapidly constructs tensor programs but also brings these programs closer to a global optimum. To achieve this, we propose Gensor, a graph-based construction tensor compilation method for deep learning. Unlike tree-based methods, Gensor models the tensor program construction into a graph traversal process. The nodes in the graph represent tensor programs, while the edges represent scheduling primitives, which are partly listed in Table \ref{tab:schedule_primitives}. Each node has multiple outgoing edges, representing possible scheduling primitive choices for the tensor program. Gensor also improves the existing scheduling primitives by supporting the virtual thread (vThread) technology. Specifically, we design an enhanced tensor program IR (ETIR) as a new representation for tensor programs. Furthermore, Gensor exploits the independent and memory-less properties of tensor programs, utilizing Markov analysis to guide the selection of scheduling primitives. In detail, Gensor uses predefined formulas to determine the probability of each scheduling primitive being selected. These formulas use the hardware platform's computing and memory architecture, along with the current tensor program's performance, as input. This way, Gensor can systematically explore the construction space by applying thoughtfully designed rules within a probabilistic framework. This process will discover the optimal combination path for optimization, thereby enhancing tensor computation performance across different hardware platforms.

Gensor improves the performance of tensor programs over existing tree-based methods while keeping the optimization time within the same order of magnitude. The slight increase in optimization time has a negligible impact on overall performance. Furthermore, in the context of dynamic DNN scenarios, the infrequent occurrence of optimization contributes a smaller proportion in comparison to long-term inference computations, further reducing its impact on performance.

\begin{table}
  \centering
  \renewcommand{\arraystretch}{1.12}
  \scriptsize
  \caption{Examples of schedule primitives in Gensor.}
  \label{tab:schedule_primitives}
  \begin{tabular}{ p{0.8cm} | p{4.5cm} | p{2cm} }
    \toprule
    \textbf{Name} & \textbf{Description} & \textbf{Formula} \\
    \hline
    split & divide a loop into several sub-loops & $L \rightarrow (L_1, L_2)$ \\
    \hline
    fuse & merge several loops into one loop & $(L_1, L_2) \rightarrow L$ \\
    \hline
    tile & split and fuse to divide multi-loops into tiles & $L \rightarrow [T_1, T_2]$ \\
    \hline
    unroll & unroll the loop & $L \rightarrow \sum_{i=1}^{n} L_i$ \\
    \hline
    cache & use multi-level cache to load/store data & $C(T)$ \\
    \bottomrule
  \end{tabular}
\end{table}

Our evaluation of operators and neural networks shows that Gensor can generate operator kernels in seconds, with performance improved by up to 30\%. It also achieves a high speedup for end-to-end models like ResNet and GPT-2.

This paper makes the following contributions:
\begin{itemize}
\item We propose Gensor, a novel graph-based construction tensor compilation method to achieve a larger construction space and higher flexibility compared to existing tree-based methods.

\item We exploit the independent and memory-less properties of tensor programs, proposing the utilization of Markov analysis to explore the construction space.

\item Extensive experiments indicate that Gensor significantly outperforms the state-of-the-art construction tensor compiling methods on GPUs for cloud servers and edge devices, with better applicability in scenarios such as dynamic DNNs.
\end{itemize}

\section{Background and Motivation}

\subsection{Background}

Tensors are multi-dimensional arrays that serve as fundamental data structures in deep learning. Deep learning frameworks like TensorFlow\cite{abadi2016tensorflow} and PyTorch\cite{paszke2019pytorch} extensively utilize tensors to represent data, including model weights, inputs and outputs. Tensor programs, which consist of a series of tensor operators, form complete operators in DNNs. The tensor compilation is a specialized technique for automatically optimizing and compiling code for tensor programs.

The tensor compiler aims to transform high-level operators represented by tensor programs into efficient machine codes that can fully leverage the capabilities of hardware's parallel architecture, such as multi-core processors and vector instructions. This process typically involves the following steps:


\textit{Optimization}: Compilers like TVM\cite{chen2018tvm} apply a suite of optimization operations known as scheduling primitives, such as loop tiling, loop unrolling, and memory access optimization, which are listed in Table \ref{tab:schedule_primitives}. Then, the compilers perform tuning by testing different combinations of scheduling primitives and various scheduling parameters, such as tiling sizes, to enhance parallelism, thus improving the memory reuse rate.

\textit{Code Generation}: After optimization, the compilers translate the optimized IR into low-level codes tailored for the specific hardware platform, called codegen. These codes could be machine codes for CPUs, CUDA\cite{nickolls2008scalable} codes for GPUs, or assembly instructions for other platforms.

The main advantage of tensor compilers is generating customized high-performance code for different hardware architectures without manual optimization. Machine learning scientists and engineers can thus focus on designing algorithms rather than worrying about the underlying hardware details.

Besides, graph traversal is crucial for efficiently navigating interconnected structures and uncovering patterns by traversing nodes and edges. Markov analysis, which captures probabilistic state transitions, offers a powerful method for investigating complex systems represented by graphs\cite{newman2005measure}.

\begin{figure}[ht]
    \centering
    \includegraphics[width=0.9\linewidth]{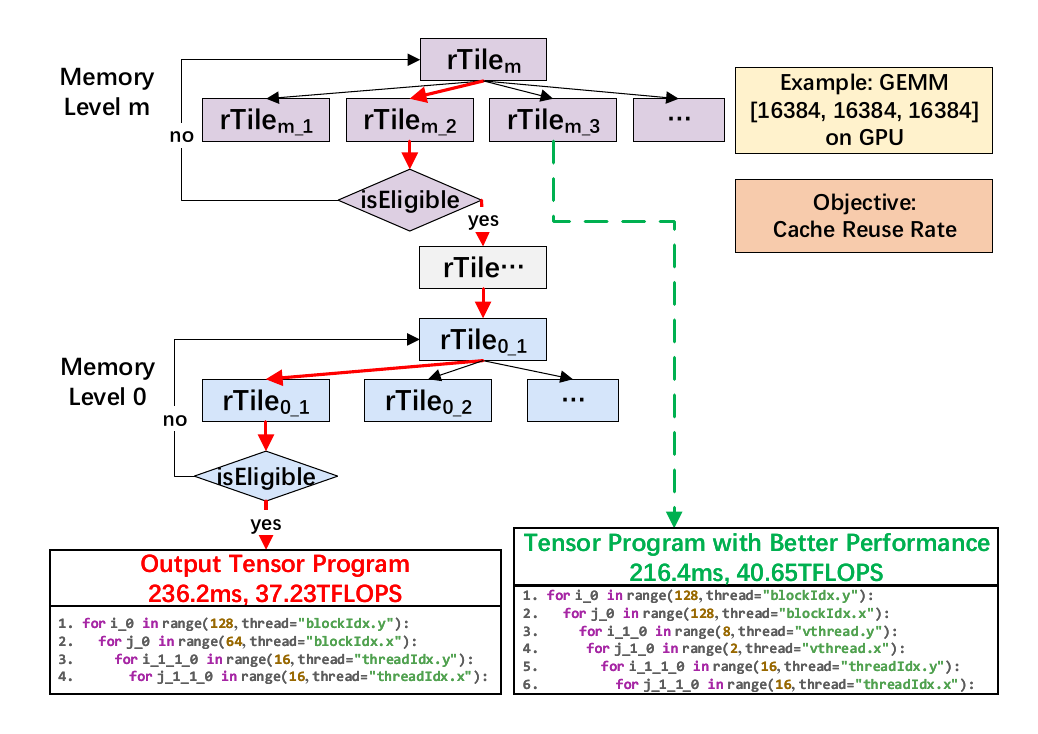}
    \caption{Unidirectional tree structure with one single objective using Roller. The red arrow indicates the solution identified by Roller. The green arrow indicates a solution with higher FLOPS (floating point operations per second), representing higher GPU throughput, namely better performance. The performance difference between the two solutions is 9\%.}
    \label{fig:roller}
\end{figure}

\subsection{Motivation}
\label{ssec:motivation}

\begin{table*}
    \scriptsize
    \centering
    \begin{tabular}{c|c|c|c|c}
    \toprule
          & \textbf{Manually-tuned} & \textbf{Searching} & \textbf{Tree-based Construction} & \textbf{Graph-based Construction}\\
    \hline
        Examples & cuBLAS & autoTVM, Ansor & Roller & \textbf{Gensor} \\
    \hline
        High-Performance & $\checkmark$ & $\checkmark$ & $\times$ & \CheckmarkBold \\
    \hline
        Agile-Development & $\times$ & $\checkmark$ & $\checkmark$ & \CheckmarkBold \\
    \hline
        Good-Flexibility & $\times$ & $\times$ & $\checkmark$ & \CheckmarkBold \\
    \bottomrule
    \end{tabular}
    \caption{Comparison of Different Tensor Program Optimization Methods. We list the performance of generated tensor programs, the agility of development, and the flexibility of the methods in order. The agility of development refers to the time and labor cost it takes to generate a tensor program each time.}
    \label{tab:comp_tensor}
\end{table*}

In practical applications like dynamic models, the structure and input shapes of operators often evolve, presenting a challenge for efficiently parallelizing tensor programs. Manual tuning, typically practical for operators with static shapes, encounters difficulties when dealing with operators featuring varying channels or input shapes\cite{han2021dynamic}. Furthermore, tensor compilation using searching methods is restricted by their inherently high computational overhead\cite{zheng2021tenset}, making it unsuitable for rapid adaptation to ever-changing models. 

Meanwhile, the construction tensor compilation based on the tree structure, such as Roller, can rapidly produce tensor programs. However, the computing performance (FLOPS) of these tensor programs is limited, which can only achieve up to 70\% of that achieved through searching approaches\cite{zhu2022roller}. For instance, when optimizing the general matrix multiplication (GEMM) using Roller, Fig.\ref{fig:roller} demonstrates that at least one optimization path outperforms the path identified by Roller.

As Fig.\ref{fig:roller} shows, the problem of the current construction methods lies in that they are based on the traversal of the tree structure. In this structure, the tree nodes represent tensor programs, while the edges represent scheduling primitives. This approach determines the unidirectional feature based on one singular objective, such as cache reuse rate, disregarding other critical performance-influencing factors like memory conflicts. Consequently, the tree structure results in a limited search space and a lack of flexibility in the optimization. Superior scheduling combinations may be overlooked when traversing upper-level nodes due to the suboptimal performance of the single objective. This results in the traversal order not being completely consistent with the performance order.




On the contrary, abstracting the tensor construction process as a graph structure can efficiently expand the search space with multiple objectives. Additionally, graph traversal supports state backtracking, improving flexibility through fast construction. Furthermore, the Markov property is beneficial in this issue because the profit of the state depends only on the current tensor program, not on the sequence of previous scheduling, thus exhibiting an independent and memory-less feature. Therefore, Gensor treats the construction graph as a state space, and then uses Markov analysis to solve it.



Table \ref{tab:comp_tensor} lists three key indicators of tensor program optimization methods and provides representative examples. The graph-based construction tensor compilation performs well in all three indicators.

\section{Overview}
\label{sec:overview}

\begin{figure}[ht]
    \centering
    \includegraphics[width=0.9\linewidth]{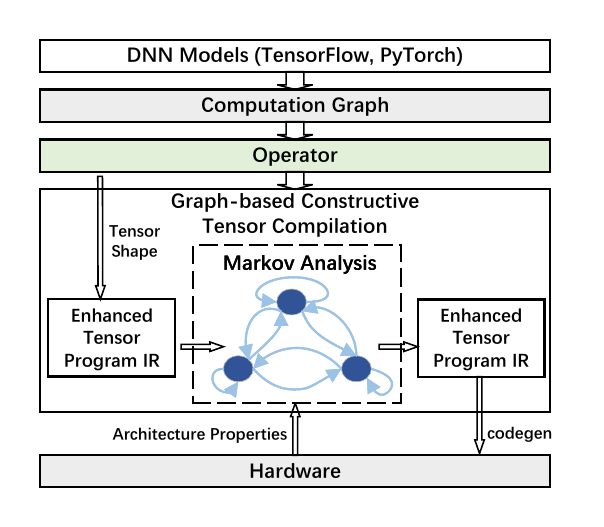}
    \caption{Overview of Gensor.}
    \label{fig:system_arch}
\end{figure}

Gensor is a graph-based construction tensor compilation method. Fig.\ref{fig:system_arch} illustrates the workflow of Gensor. Gensor abstracts the construction space into a graph model. The input of Gensor is the DNN's operators, which will be represented as our proposed ETIR. Then, Gensor models ETIR as nodes and scheduling primitives as edges. Therefore, the process of tensor program construction optimization is abstracted as a graph traversal process. To explore the construction graph, Gensor utilizes Markov analysis with the guide of the hardware information. Finally, Gensor translates the obtained high-performance ETIR into target codes.

As depicted in Fig.\ref{fig:system_arch}, Gensor resides between the tensor program layer and the hardware layer. In implementation, Gensor is built based on TVM and Roller. Gensor uses Markov analysis to determine the optimal configuration of each variable within the ETIR, such as loop lengths. The nodes in the graph represent the states in Markov analysis, and the edges represent the actions transitioning between states with certain probabilities. In detail, state transitions are driven by scheduling primitives, including loop tiling, caching and setting virtual threads. The probabilities of state transitions are determined by the normalized performance improvement of the tensor program resulting from the scheduling action. Meanwhile, the probabilities are also guided by the architecture of the target hardware, represented by computing and memory features. By employing the scheduling primitive-based graph space, Gensor can effectively traverse the expansive state space to optimize tensor programs.


To better express and utilize the thread-level parallelism of the hardware, ETIR incorporates virtual thread features to allow more granular control over tensor programs. The virtual thread is a logical unit of execution of a parallel operation within a grid of threads. Through a refined tiling strategy, ETIR can mitigate data transmission bus conflicts, thereby efficiently utilizing cores for concurrent data processing. This feature empowers ETIR with increased expressiveness in generating high-performance tensor programs.

The selection process includes a series of transitions guided by probabilistic rules based on the hardware architecture and the current tensor program's performance. This ensures a systematic and global optimization process. The hardware architecture properties contain peak computing performance, memory hierarchies, and parallelism features. This hardware-aware approach ensures that the generated tensor programs are well-suited to the physical constraints of the target hardware.

The probabilistic rules form a state transition probability matrix, enabling the compiler to navigate different optimization paths. By leveraging this matrix, the compiler can select the optimization path that promises the highest expected efficiency without repeatedly iterating code generation and profiling on the actual hardware during the optimization process. Consequently, Gensor will obtain high-performance tensor programs tailored to the specific hardware architecture.

\section{Method}

\begin{figure*}[ht]
    \centering
    \includegraphics[width=0.92\linewidth]{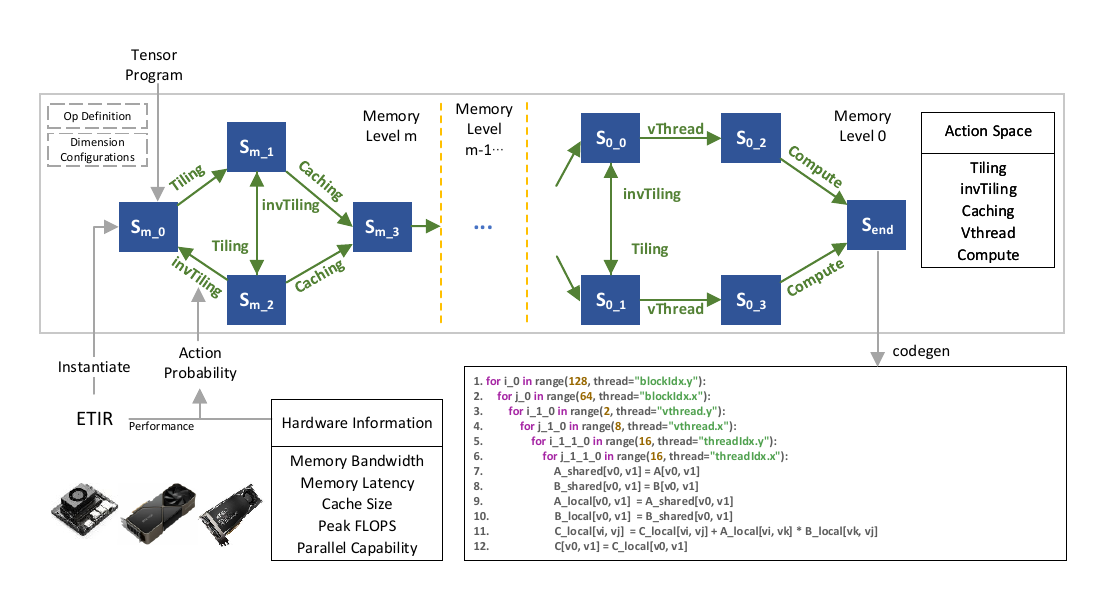}
    \caption{Illustration of Gensor. The blue blocks represent the nodes, namely the possible tensor programs in the construction graph. The green arrows represent the edges, namely the possible scheduling primitives in the construction graph. The \textit{Memory Level} means the order of cache levels in the target hardware. A higher level means the memory is closer to computing units.}  
    \label{fig:algo_arch}
\end{figure*}

The graph structure is effective in expressing and managing complex optimizing combinations. We employ Markov analysis to better explore the construction space, thus uncovering better combinations for parallel computing optimization. The Markov analysis method excels at exploiting the independent and memory-less properties of tensor programs, making it highly suitable for exploring the construction space of tensor programs. The following discussion will delve into the states, actions, and traversal strategies within the context of Gensor's graph architecture. Fig.\ref{fig:algo_arch} shows the architecture of Gensor.



\subsection{States and Actions}
\textbf{\textit{States}}: 
In traditional tile-based IR, the tensor program is divided into small blocks known as tiles. These tiles are allocated to physical memory and scheduled according to specific rules. Nevertheless, this tile-based IR lacks the ability to thread scheduling with finer granularity. Hence, ETIR integrates virtual threads for parallel computing and thread-level optimization, building upon the existing tile-based tensor IR. ETIR builds upon the existing tile-based IR. The data structure of ETIR is shown below.

\begin{lstlisting}[language=C++, numbers=left, xleftmargin=10pt, numbersep=5pt, tabsize=4, breaklines=true, basicstyle=\footnotesize]
class ETIR()
{
  TensorIR tir;      // representation of tensor program
  Axis axis;         // spacial and reduce axis
  Shape shape;       // shapes of tensor program

  int numLevel;      // memory levels
  int curMemLevel;   // current scheduling memory level 
  List etiles;       // tiles of each memory level
  List evThreads;    // virtual thread configuration

  void tile();       // tiling 
  void invTile();    // inverse tiling 
  void setVThread(); // set virtual threads
  void cache();      // switch memory level to the next
};
\end{lstlisting}

    
States, instantiated with ETIR, represent the nodes of the construction graph. The state specifically includes the following major components:
\textit{Tensor Tiling} (etiles): States describe how tensor programs are divided into different tiles while specifying the position and size of each tile. This information determines the parallel execution configurations of tensor programs on the hardware. Besides, the memory level refers to the hierarchy of storage types, from fast, small-capacity caches to slower, larger-capacity main memory and secondary storage.


\textit{Thread status} (evThreads): States include the current execution status of each virtual thread. This information is beneficial for load balancing and parallelism.

\begin{figure}[ht]
    \centering
    \includegraphics[width=1\linewidth]{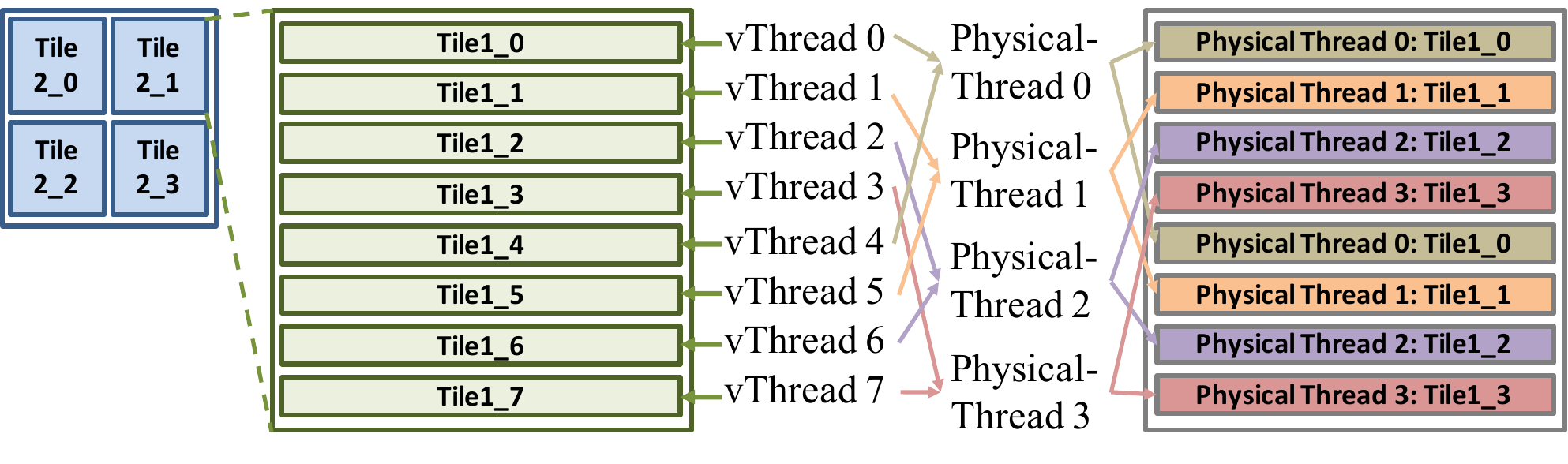}
    \caption{Diagram of virtual threads in ETIR.}
    \label{fig:etir}
\end{figure}

VThread technology is commonly used in manual optimization to optimize the spacing and synchronization between threads\cite{yoon2016virtual}. It abstracts computing tasks into independent execution units, allowing programs to logically create more threads than physical cores. As shown in Fig.\ref{fig:etir}, each tile can be further divided into multiple virtual threads in ETIR. In the code generation phase, it will be reaggregated into actual physical threads. Gensor incorporates vThread action into the auto-scheduling space. The motivation is to leverage the performance benefits brought by vThread, achieving automatically assigning tasks to virtual threads and fine-grained control over resource access. By adjusting vThread configurations, computing workloads are more evenly distributed, reducing memory bus conflicts and increasing cache efficiency. Therefore, we add the \textit{setVthread()} primitive in ETIR to provide a flexible mechanism for thread-level optimization.


For instance, a state can describe the specific allocation of tensor programs to hardware resources, with each resource associated with a particular tile. The state contains information about the current resource usage of each tile and the corresponding performance metrics. By defining the states this way, we can effectively capture the system's configuration and performance characteristics at each step of Markov analysis, thereby enhancing the performance of the graph traversal.

\textbf{\textit{Actions}}: 
The edges in the construction graph are abstracted into actions, representing the different possible scheduling primitives taken by each state. The actions make the transition from one state to another, thereby impacting the performance of the tensor program. The actions specifically include the following three components. 
\textit{Tile partitioning and scheduling}: This action divides tensor programs into smaller tiles or reschedules allocated tiles to optimize parallelism and load balancing. These operations are based on the dependencies between tiles and the usage of hardware resources. 
\textit{Caching for Input/Output}: This action changes the memory level to be accessed based on the current memory level's utilization to improve data reading and writing speed. 
\textit{Thread-level optimization}: This action assigns tensor programs to different vThreads to achieve better load balancing and parallelism.

\begin{figure*}[ht]
    \centering
    \includegraphics[width=0.9\linewidth]{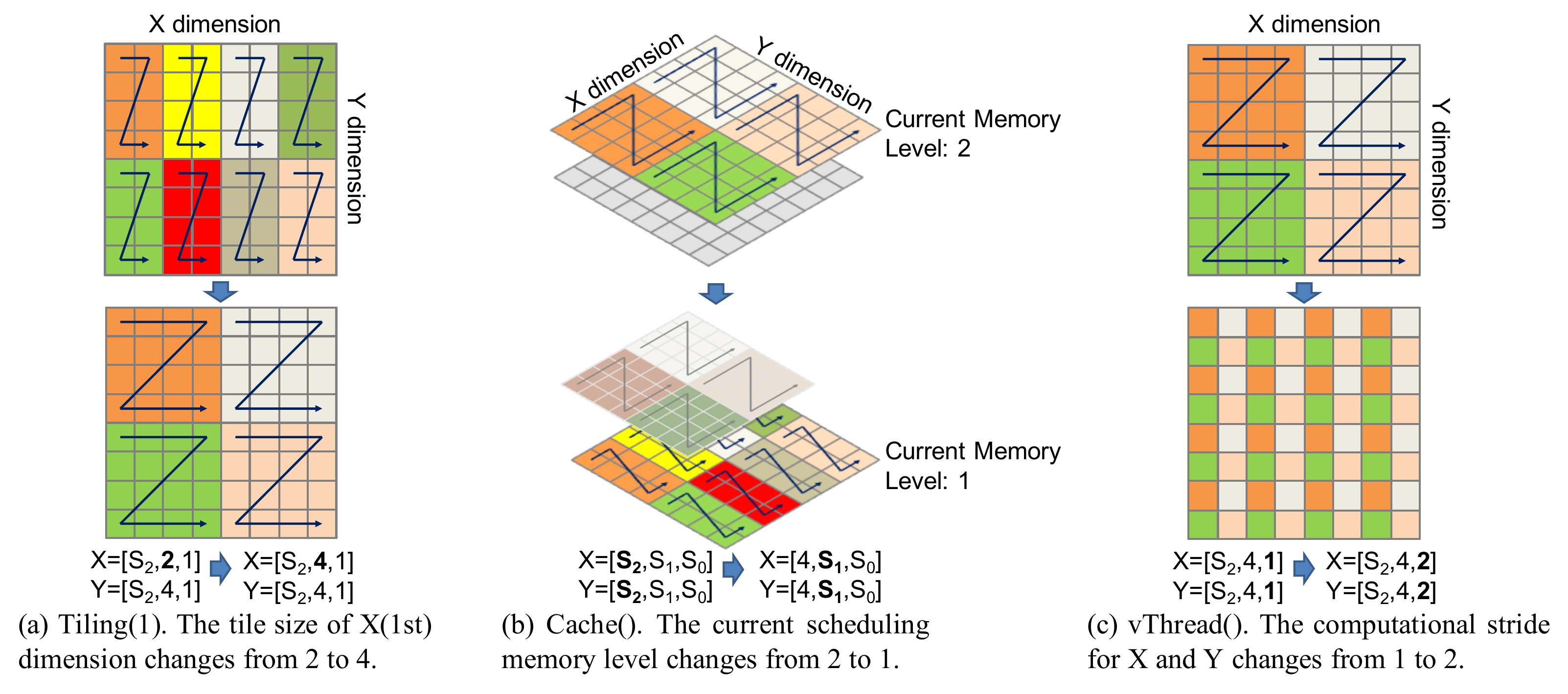}
    \caption{Illustration of Actions. Each color represents a tile corresponding to the elements that a thread needs to compute.}  
    \label{fig:actions}
\end{figure*}

\subsection{Transition Probability}
\label{ssec:trans_prob}

We calculate the benefits of the actions using thoughtfully designed formulas. These formulas are jointly defined by the computing and memory performance of the current tensor program and the hardware architecture. Then, we normalize them to represent the probabilities of state transition.

\textbf{\textit{Tiling}}: 
Tiling for nested loops enhances data locality by breaking down large loops into smaller, more manageable blocks. This action reduces memory traffic, allowing for efficient use of hierarchical caching. The tiling/invTiling action is accomplished by increasing or decreasing the tile size of each dimension, as illustrated in Fig.\ref{fig:actions}(a).

Formula \ref{eq:tiling} defines the benefit from the tiling action. The formula balances the reduction in memory traffic against the increase in memory footprint resulting from tiling. Here, $Q(T)$ and $Q(T')$ represent the memory traffic before and after tiling, respectively, while $F(T)$ and $F(T')$ represent the memory footprint. The numerator, $\frac{Q(T)}{Q(T')}$, indicates the decrease in memory traffic. The denominator, $\frac{F(T)}{F(T')}$, reflects the increase in memory footprint. A higher ratio indicates a higher memory reuse rate, meaning a higher probability of being selected.

\begin{equation}
\label{eq:tiling}
\textit{Benefit}_{Tiling} = \frac{\frac{Q(T)}{Q(T')}}{\frac{F(T)}{F(T')}} = \frac{{Q(T)F(T')}}{Q(T')F(T)}
\end{equation}

\textbf{\textit{Caching}}: 
Caching for input and output data optimizes computation by loading and offloading data into and from faster, higher-level memory, which is closer to the computing units. The cache action means switching the current scheduling memory level to the next, as illustrated in Fig.\ref{fig:actions}(b).

Formula \ref{eq:caching} defines the benefit from the caching action. The formula quantifies the performance improvement from caching by comparing the access latency and bandwidth of different memory levels. Here, $L$ represents the latency of the memory, $S$ represents the size of data exchanged, and $B$ represents the bandwidth of the memory. Specifically, it calculates the relative speed increase in data access when leveraging caches. A higher ratio indicates a shorter reading and writing time, meaning a higher probability of being selected. 

\begin{equation}\label{eq:caching}
\textit{Benefit}_{Caching} = \frac{L_{low} + \frac{S_{data}}{B_{low}}}{L_{high} + \frac{S_{data}}{B_{high}}}
\end{equation}

\textbf{\textit{Setting virtual threads}}: 
This action increases computing parallelism by interleaving tasks across physical threads, thereby reducing resource contention. This enables adaptable and efficient utilization of hardware resources in multi-threads environments. The action refers to the division of memory data accessed by physical-threads, as illustrated in Fig.\ref{fig:actions}(c).

Formula \ref{eq:vthread} defines the benefit from implementing vThread action. The formula integrates virtual threads (V) into a parallel computing model, calculated as the ratio of original bank conflicts to those with virtual threads. Here, \textit{x} and \textit{y} represent the width and height of the tile processed in parallel, while W denotes the bank width and V represents the number of virtual threads. The ceiling operations ensure that even partial overlaps in bank accesses are accounted for. The formula estimates potential reductions in bank conflicts by distributing memory accesses across different banks.

\begin{equation}\label{eq:vthread}
\textit{Benefit}_{vThread} = \frac{\left\lceil \frac{x}{W} \right\rceil \cdot y}{\left\lceil \frac{x}{V \cdot W} \right\rceil \cdot y} = \frac{\left\lceil \frac{x}{W} \right\rceil}{\left\lceil \frac{x}{V \cdot W} \right\rceil}
\end{equation}

\subsection{Transition Policy}



The Algorithm \ref{alg:GetOptimalTP} describes the procedure for optimizing tensor programs using a simulated annealing approach based on the graph structure. It begins by initializing a temperature variable, \texttt{T}, as well as an initial ETIR, \texttt{e}, with the provided tensor shape and dimension configurations. The algorithm iterates in a loop until $T$ falls below a predefined threshold. Within this loop, the actions and dimension configurations are reset. Then, a programming policy, \texttt{getProgPolicy()}, is invoked to generate a new action, \texttt{ac}, and dimension configurations, \texttt{dim\_configs}, based on the probability of each action, which is normalized by its benefit. The policy \texttt{getProgPolicy()} is detailed in Algorithm \ref{alg:GetSP}. Since each action's benefit is represented by its acceleration ratio, Gensor sums the benefits as the denominator, with each individual benefit serving as the numerator. The algorithm first gets the transition probability of each action. Then, it utilizes a roulette selection to select the scheduling primitive. A higher probability indicates that the action is more likely to be selected. The probability calculation for each action is independent, as it considers only the theoretical acceleration benefits of that specific action. Additionally, in the formula of each action, the parameters related to cache size and bandwidth are the theoretical performance metrics.

An annealing algorithm is employed to ensure that the traversal converges to the next level of cache and ultimately terminates. As the temperature falls, the probability of selecting the cache action increases by multiplying the transition probability by \(\frac{3}{1+e^{-\frac{\ln(5)}{10}(t-10)}}\). This ensures the convergence of the results. Besides, Gensor conducts memory check for each transition. If memory required for the configuration exceeds the cache capacity, the probability is directly set to 0.

The chosen primitive is then applied to the current ETIR to derive a modified ETIR, denoted as $e'$. The new ETIR is appended to the list of top results with a probability of $1-\frac{1}{1+e^{-0.5(-\log{t}-10)}}$. The temperature is halved at each iteration to gradually reduce the probability of remaining at the same memory level, thereby transitioning to higher level memory, and finally converging to an optimal solution. The process ensures that a diverse set of tensor program configurations is explored, with more effective tensor programs yielded.

\begin{algorithm}
    \caption{Construction Process of Gensor.}
    \label{alg:GetOptimalTP}
    \begin{small}
    \begin{algorithmic}[1] 
        \STATE \textbf{Input:} {T:\ temperature, threshold, op, tensor\_shape, dim\_configs}
        \STATE \textbf{Output:} {optimal tensor programs}
        \STATE $e \gets \text{ETIR} (tensor\_shape, dim\_configs)$
        \WHILE {$T > threshold$}
            \STATE {$ac, dims \gets \text{getProgPolicy}(e, T)$}
            \STATE {$e' \gets \text{applyProgPolicy}(e, ac, dims)$}
            \IF {$rand() < 1-\frac{1}{1+e^{-0.5(-\log{T}-10)}}$}
                \STATE {$top\_results.append(e')$}
            \ENDIF
            \STATE {$e \gets e'$}
            \STATE {$T \gets T / 2$}
        \ENDWHILE
        \RETURN {$e$}
    \end{algorithmic}
    \end{small}
\end{algorithm}

In detail, Gensor represents the memory tiling configuration of each loop (dimension) of the tensor program as \(D\)=[\(T_L\),...,\(T_1\),\(T_0\)], where L denotes the layer number of cache (in Nvidia GPU, L=2), and T represents the size of each tile. \(T_0\) indicates the computational stride for each thread (i.e. virtual-thread). For the current state, Algorithm \ref{alg:GetOptimalTP} calls Algorithm \ref{alg:GetSP} to calculate the benefit for each action. Then the benefits of actions are normalized as their probabilities. The action is selected based on these probabilities. Then the state transitions to the next state until the process converges.

\begin{algorithm}
    \caption{Get Scheduling Policy with Markov analysis.}
    \label{alg:GetSP} 
    \begin{small}
    \begin{algorithmic}[1]
    \STATE \textbf{Input:} {e:\ ETIR, T:\ temperature} 
    \STATE {probList = [] \textit{// Initialize Probability List}}
    \STATE {actionMap = \{\} \textit{// Initialize Action Map}}
        \FOR{$ac\ \text{from}\ 0\ \text{to}\ n$}
            \FOR{$d\ \text{from}\ 0\ \text{to}\ dims$}
                \STATE {$prob\ \gets\ \text{getBenefit}(e, ac, [d], T))$}
                \STATE {$probList.append(prob)$}
                \STATE {$actionMap[ac].append([d])$}
            \ENDFOR
        \ENDFOR
        \STATE {$probList \gets \text{Normalize}(probList)$}
        \FOR{$i\ \text{in}\ probList$}
            \IF{$rand() >= probList[i]$}
                \STATE {$ac, dim \gets \text{getAction}(ProbList, actionMap)$}
                \RETURN {$ac, dim$}
            \ENDIF
        \ENDFOR
    \end{algorithmic}
    \end{small}
\end{algorithm}

\subsection{Convergence and Validity Analysis}
We examine the Markov chain characterized by a finite state space $S$ and a transition probability matrix $P$. The states in $S$ represent ETIR, which have a finite number. $T(d,l)$ in S describes the tile size T of dimension $d$ on the cache level $l$. $P(i,j)$ describes the probabilities of actions transitioning from state $i$ to state $j$. First, we ensure that the Markov chain is irreducible. Mathematically, the chain is irreducible if, 
$\forall i, j \in S, \exists n > 0 : P^n(i, j) > 0$. In Gensor, since we set up the inverse tiling action, the states within the same-level memories can be converted to each other, thus ensuring the irreducibility of the same-level memories.

Next, we verify aperiodicity, which requires any state $i$ to satisfy $gcd\{ n \in \mathbb{N} : P^n(i, i) > 0 \} = 1$. According to Algorithm \ref{alg:GetOptimalTP} and the Fig.\ref{fig:algo_arch}, the number of steps for a state to return to itself may be 2, 3, or others, and the \textit{gcd} of these is 1.
In summary, by demonstrating irreducibility and aperiodicity, we prove that the Markov process for construction tensor compilation is convergent. Consequently, we can compute a stationary distribution. This demonstrates the effectiveness of Markov analysis method in solving the graph-based construction tensor compilation problem.


{
\noindent
\begin{equation}\label{eq:StateMatrix}
\scriptsize
S=\begin{bmatrix}
T_{1,L} & \dots & T_{1,1} & T_{1,0} \\
T_{2,L} & \dots & T_{2,1} & T_{2,0} \\
\vdots & \vdots & \ddots & \vdots \\
T_{D,L} & \dots & T_{D,1} & T_{D,0} \\
\end{bmatrix} 
, P=\begin{bmatrix}
P_{1,1} & P_{1,2} & \dots & P_{1,n} \\
P_{2,1} & P_{2,2} & \dots & P_{2,n} \\
\vdots & \vdots & \ddots & \vdots \\
P_{n,1} & P_{n,2} & \dots & P_{n,n}
\end{bmatrix}
\end{equation}
}

\noindent
where
$P(i,j) = Norm(Benefit(Action(i\rightarrow j))) \geq 0$ and $\sum_{j} P(i,j) = 1$

Next, we aim to prove that the convergence state of this process is the state with the maximum payoff. In the given Markov process, the payoff for each action is defined by the ratio of benefit associated with the states before and after the action. It suggests that the payoff depends on multiplicative effects rather than additive ones. Therefore, we adjust the payoff equation from a summation to a product form. This ensures that the payoff calculation accurately captures the compound effect of state transitions on the action's overall payoff. The initial state refers to the unscheduled state without partitioning, caching, or virtual threads.

We denote $V(i)$ as the value function for state $i$ (denoted by $s_{i}$), representing the maximum expected payoff starting from $s_{i}$. The adjusted Bellman equation for $V(i)$ is:
\begin{equation}\label{eq:Vi}
V(i) = \max_{a \in A(i)} \left( \pi(a \mid i) V(j) \right)
\end{equation}
where the benefit associated with each action $a$ at state $i$ is defined to be the normalized probability of choosing that action in Gensor, i.e., $\pi(a|i)=Normalized(Benefit_{a}(i))$, where $\pi(a|i)$ is the probability of taking action $a$ at state $i$.
The policy $\pi$ selects the action $a$ at each state $i$ that maximizes $\pi(a|i)V(j)$. Consequently, $V(i)$ reflects the maximum benefit achievable at that state under optimal action choices.

For state $i$, at the $(k+1)$-th iteration, the value of the state is updated as follows:
\begin{equation}\label{eq:ViK+1}
V_{k+1}(i) = \max_{a \in A(i)} \left( \pi(a \mid i) \cdot V_k(j) \right)
\end{equation}
where $j$ is the state to which the system transitions after taking action $a$ from state $i$.

Repeat the steps for the states until the value of each state stabilizes, meaning that $V_{k+1}(i) = V_k(i)$ is satisfied for all $i$.
Since the state space is finite and each iteration selects the optimal action, this guarantees the monotonic non-decreasing nature of the benefits. Furthermore, this ensures that the algorithm will converge to a fixed point $s_{i^*}$ within a finite number of steps, which is the benefits value for each state.




Therefore, the convergence state $i^*$ in the Markov process is the state with the maximum payoff. The policy $\pi$, which selects the action $a$ maximizing $V$ at each state, is thus valid. According to practical applications, convergence can generally be achieved after about 100 iterations, reflecting the optimization speed of Gensor.


\section{Evaluation}

We implement Gensor in Python and use TVM for code generation. We verify Gensor on GPUs for both cloud servers and edge devices. The configurations of the experimental environment are shown in Table \ref{tab:ex_setup}. 

\begin{table}[htbp]
\centering
\caption{Experimental environment setup.}
\scriptsize
\label{tab:ex_setup}
\renewcommand{\arraystretch}{1.12} 
\begin{tabular}{c|c|c}
\toprule
\textbf{Metric} & \textbf{Edge Device}  & \textbf{Cloud Server}\\
\hline
CPU &  ARM v8l r1   & Intel i7-13700K \\
GPU & NVIDIA Orin Nano & NVIDIA RTX 4090 \\
GPU Memory Size & 8GB & 24GB \\
GPU Power & 15W & 450W \\
\bottomrule
\end{tabular}
\end{table}

\begin{table*}[htbp]
\centering
\caption{A subset of operator configurations in the benchmark.}
\scriptsize
\label{tab:op_example}
\begin{tabular}{c|c|l|c}
\toprule
\textbf{Op} & \textbf{Formula} & \textbf{\ \ \ \ \ \ \ \ \ \ \ \ \ \ \ \ \ \ \ \ \ Shape} & \textbf{Label} \\
\hline
  &  & I=[128,256,30,30],K=[256,256,3,3],S=2 & C1 \\
Conv2d & $O=(I * K) \downarrow_S$ & I=[128,128,28,28],K=[128,128,3,3],S=1 & C2 \\
  &  & I=[128,128,58,58],K=[128,128,3,3],S=2 & C3 \\
\hline
 & $C_{ij}=\alpha \sum_{k=1}^{K} (A_{ik}$ & MKN=[8192,8192,8192] & M1 \\
GEMM & $ B_{kj})+\beta C_{ij}$ & MKN=[65536,4,1024] & M2 \\
 & $i\in[0,M),j\in[0,N)$ & MKN=[65536,1024,4096] & M3 \\
\hline
 & $y_i=\alpha \sum_{n=1}^{N}$ & MN=[16384,16384] & V1 \\
GEMV & $ (A_{in}x_n)+\beta y_i$ & MN=[16384,8192] & V2 \\
 & $i\in[0,M)$ & MN=[16384,1000] & V3 \\
\hline
 & $O_{x,y}=\frac{1}{F^2}  $ & I=[16,48,48,48], F=2, S=2 & P1 \\
AvgPooling2d & $ \sum_{i=1}^{F} \sum_{j=1}^{F}$ & I=[128,168,83,83], F=2, S=2 & P2 \\
 & $ I_{S \cdot x + i, S \cdot y + j} $ & I=[128,617,21,21], F=3, S=2 & P3 \\
\bottomrule
\end{tabular}
\end{table*}

\begin{figure*}[ht]
    \centering
    \includegraphics[width=0.9\linewidth]{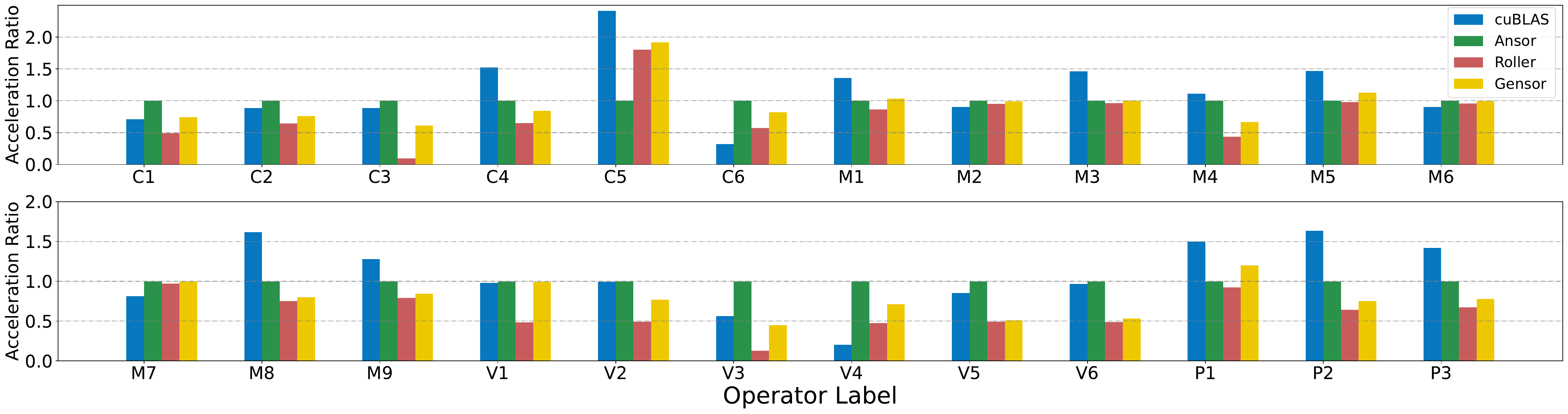}
    \caption{Operator performance on the RTX 4090 GPU. The horizontal axis represents the operator label listed in Table \ref{tab:op_example}, and the vertical axis represents the relative performance (FLOPS) compared to Ansor.}
    \label{fig:operator_performance_4090}
\end{figure*}

\begin{figure*}[ht]
    \centering
    \includegraphics[width=0.9\linewidth]{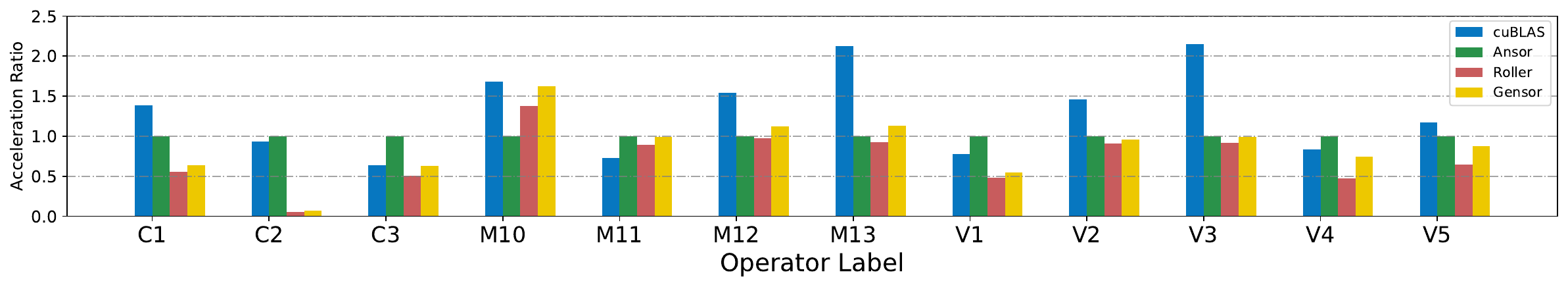}
    \caption{Operator performance on the Orin Nano GPU compared to Ansor. The horizontal axis represents the operator label listed in Table \ref{tab:op_example}, and the vertical axis represents the relative performance (FLOPS) compared to Ansor.}
    \label{fig:operator_performance_ORIN}
\end{figure*}

To evaluate our approach, we complete three sets of experiments. The first set in subsection \ref{ssec:op_perf} tests the performance of operators to verify Gensor's High-Performance feature in Table \ref{tab:comp_tensor}. The second set in subsection \ref{ssec:comp_time} measures the time required for tensor compilation to verify Gensor's Agile-Development feature. The third set in subsection \ref{ssec:ee_perf} evaluates the effect of Gensor on optimizing the end-to-end neural networks, including the models containing operators with dynamic shapes. This verifies Gensor's Good-Flexibility feature. These experiments compare the performance of Gensor with other representative tensor compilation methods.



\subsection{Operator Performance}
\label{ssec:op_perf}


We conduct performance evaluations on a suite of 32 operator configurations with diverse shapes, including convolution, GEMM, GEMV (General Matrix Vector Multiply), and average pooling to prove Gensors's high-performance feature. These operators are widely used in DNNs. The specifics of them are detailed in Table \ref{tab:op_example}.
Our comparison methods include handwritten libraries (cuBLAS), searching tensor compilation methods, exemplified by Ansor, and the tree-based construction tensor compilation method, Roller. The outcomes of these evaluations are depicted in Fig.\ref{fig:operator_performance_4090} and Fig.\ref{fig:operator_performance_ORIN}.

The results reveal that Gensor, while ensuring the correctness of calculation, outperforms Roller on the operators' performance, achieving up to an 18\% increase on average. The highest performance improvement reaches 30\%. While gains in GEMV are noticeable (almost 1.3x faster) and modest in avgPooling (nearly 1.05x faster), Gensor's consistent lead across diverse operators underscores its robustness. Moreover, Gensor outperforms Ansor in some instances, like C5 and M1, reaching up to 1.9x of Ansor. Gensor can perform better than cuBLAS in a few cases, like M7, reaching up to 1.4x of cuBLAS. In most other cases, Gensor's performance is comparable to cuBLAS and Ansor's. Gensor can achieve 81.2\% of the performance of cuBLAS on average.

For simple or standardized operators, in which  tensor dimension sizes are close or follow a simple ratio, handtuning methods may suffice. However, for operators with highly unbalanced dimensions, where the data has a complex structure (unpredictable), or some dimensions are significantly larger or smaller than others, Gensor's performance reaches or even exceeds that of Ansor or cuBLAS. Table \ref{tab:op_metrics} shows Gensor's high-performance results in the unbalanced dimensions. This kind of shapes is quite common, especially in LLM\cite{liu2024resource}. CuBLAS, based on handcrafted templates, cannot provide specific optimizations for these special cases.

In this scenario, one dimension has a relatively small length and quickly reaches its boundary, whereas other dimensions remain far from meeting theirs. This significant disparity in the rates at which different dimensions approach their boundaries creates a complex and unbalanced search space. Therefore, heuristic-based search methods such as Ansor are more likely to produce incorrect solutions. This results in more erroneous solutions and a failure to find high-performance solutions in a fixed number of search iterations. In contrast, owing to its graph structure, Gensor can backtrack when it encounters boundary conditions. This capability enables more flexible traversal for unusual or non-standard memory access patterns and data layouts, thereby enhancing operator performance.

Table \ref{tab:ablation} presents the results of an ablation experiment evaluating the impact of graph-based construction and vThread on the performance of optimization methods. The results indicate significant improvements in FLOPS and the computing units' occupancy (SM Occ.) with the integration of graph-based construction and vThread technology. Specifically, the baseline method, Roller, exhibits the lowest FLOPS and SM occupancy. Introducing graph-based construction (without vThread) results in a noticeable performance increase, enhancing both FLOPS and SM occupancy. The addition of vThread (Gensor) further improves these metrics to a small extent. Thus, we affirm the pivotal role of graph-based construction in the method, accounting for 79.24\%, while vThread accounting for 20.76\%. Additionally, we demonstrate the valuable contribution of vThread in improving performance metrics.

In terms of memory overhead, the additional CPU memory for storing intermediate states is tens of megabytes. For GEMM with dimensions [16384,16384,16384], Roller’s maximum memory usage is 547MB, while Gensor’s is 627MB, which has minimal impact on the overall memory usage.


\begin{table}[htbp]
\centering
\caption{The hardware metric breakdown between Gensor and other methods for GEMM on the RTX 4090 GPU.}
\tiny
\label{tab:op_metrics}
\renewcommand{\arraystretch}{1} 
\begin{tabular}{p{1.3cm}|p{0.45cm}|p{0.45cm}|p{0.45cm}|p{0.45cm}|p{0.45cm}|p{0.45cm}|p{0.45cm}|p{0.45cm}}
\toprule
\multicolumn{1}{c|}{Metric} & \multicolumn{2}{c|}{Compute} & \multicolumn{2}{c|}{Memory} & \multicolumn{2}{c|}{L2 Cache} & \multicolumn{2}{c}{Execution Time} \\ 
\multicolumn{1}{c|}{} & \multicolumn{2}{c|}{Throughput} & \multicolumn{2}{c|}{Busy} & \multicolumn{2}{c|}{Hit Rate} & \multicolumn{2}{c}{(ms)} \\ 
\hline
\ \ \ \ \ \ \ Method & Gensor & Ansor & Gensor & Ansor & Gensor & Ansor & Gensor & Ansor\\
\hline
[65536,4,1024] & 18.9\% & 17.1\% & 50.9\% & 46.7\% & 99.6\% & 92.7\% & 0.287 & 0.303\\
\hline
[32768,64,2048] & 83.9\% & 76.3\% & 64.1\% & 61.7\% & 66.5\% & 51.7\% & 0.369 & 0.387\\
\hline
[16384,32,1024] & 69.2\% & 61.2\% & 82.1\% & 80.3\% & 99.2\% & 95.1\% & 0.083 & 0.091\\
\bottomrule
\end{tabular}
\end{table}

\begin{table*}[htbp]
\centering
\caption{The impact of graph-based construction and vThread on optimization methods on the RTX 4090 GPU.}
\renewcommand{\arraystretch}{1.12} 
\scriptsize
\label{tab:ablation}
\renewcommand{\arraystretch}{1} 
\begin{tabular}{c|c|c|c|c|c|c|c|c|c|c|c|c}
\toprule
\multicolumn{1}{c|}{} & \multicolumn{3}{c|}{Conv2d (C1)} & \multicolumn{3}{c|}{GEMM (G1)} & \multicolumn{3}{c|}{GEMV (V1)} & \multicolumn{3}{c}{AvgPooling2d (P1)} \\ \hline
\diagbox{Methods}{Metrics} & FLOPS & SM  Occ. & MemBusy & FLOPS & SM Occ. & MemBusy & FLOPS & SM Occ. & MemBusy & FLOPS & SM Occ. & MemBusy \\ \hline
Roller & 22.76T & 46.36\% & 21.42\% & 37.6T & 59.50\% & 42.80\% & 0.23T & 18.24\% & 3.67\% & 0.07T & 1.74\% & 26.98\% \\ \hline
Gensor w/o vThread & 31.93T & 51.92\% & 24.77\% & 43.1T & 63.81\% & 60.12\% & 0.39T & 22.13\% & 23.96\% & 0.08T & 3.08\% & 36.88\% \\ \hline
\textbf{Gensor} & \textbf{34.54T} & \textbf{52.17\%} & \textbf{25.48\%} & \textbf{45.2T} & \textbf{64.11\%} & \textbf{64.06\%} & \textbf{0.47T} & \textbf{24.24\%} & \textbf{24.80\%} & \textbf{0.08T} & \textbf{3.44\%} & \textbf{42.90\%} \\
\bottomrule
\end{tabular}
\end{table*}

\subsection{Compilation Time}
\label{ssec:comp_time}

\begin{figure}[ht]
    \centering
    \includegraphics[width=0.9\linewidth]{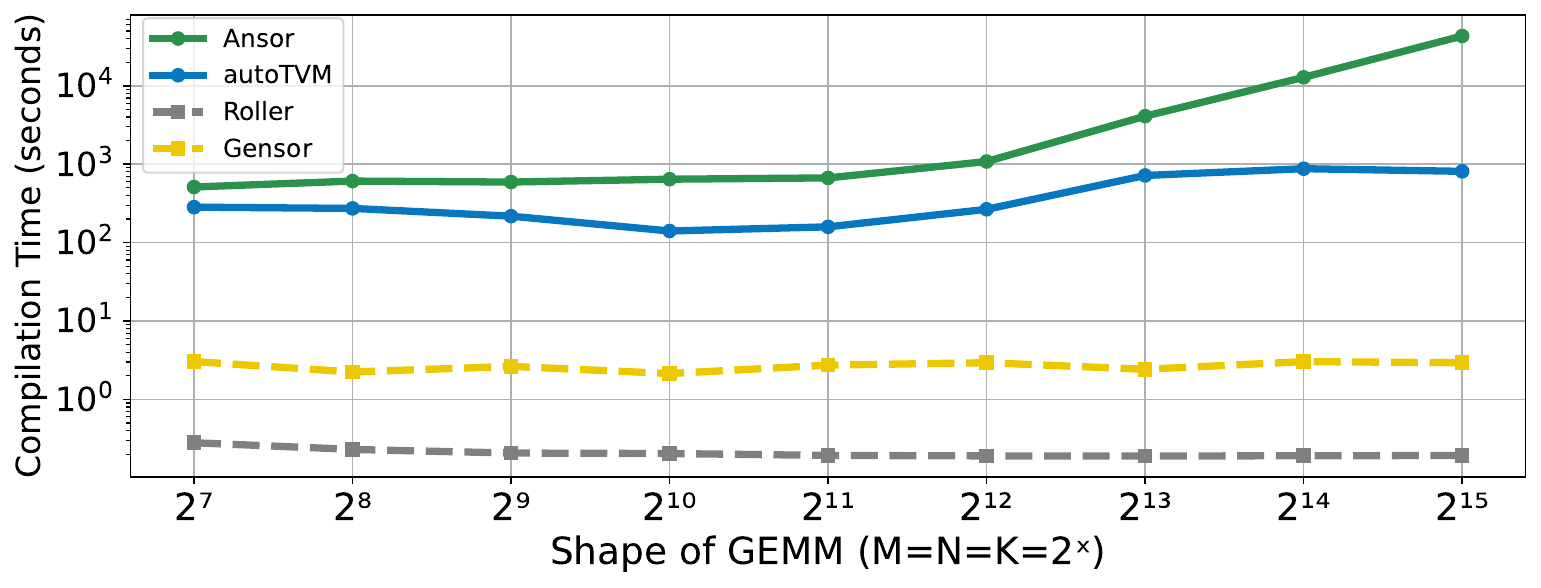}
    \caption{Compilation time for different shapes of GEMM. Gensor is approximately one order of magnitude slower than Roller and outperforms Ansor by three to five orders of magnitude, where Gensor takes a few seconds, Roller is below 1 second, and Ansor takes about 1000 seconds.}
    \label{fig:compile_time}
\end{figure}

We measure the compilation time of GEMM with different shapes and different methods. The outcomes shown in Fig.\ref{fig:compile_time} indicate that Gensor is slightly slower than Roller, the tree-based construction approach, with a time lag ranging from several hundred milliseconds to a few seconds. This is because Markov analysis involves stochastic selection and probability calculations at each step. The graph traversal is computationally more complex than deterministic tree traversal. However, this increase in compilation time is negligible because low-frequency construction accounts for a smaller proportion than long-term intelligent inference computing.

Compared to the time consumed by Ansor, Gensor outperforms Ansor by three to five orders of magnitude, showing a significant speed advantage. This suggests that although Gensor may exhibit a slight delay during the compilation phase, the overall efficiency of the optimization process is substantially enhanced by the construction method. This finding indicates that in practical applications, Gensor has the potential to deliver faster and higher optimization performance.

\subsection{End-to-end Model Performance}

\begin{figure}[ht]
    \centering
    \begin{subfigure}[b]{0.85\linewidth}
        \centering
        \includegraphics[width=\linewidth]{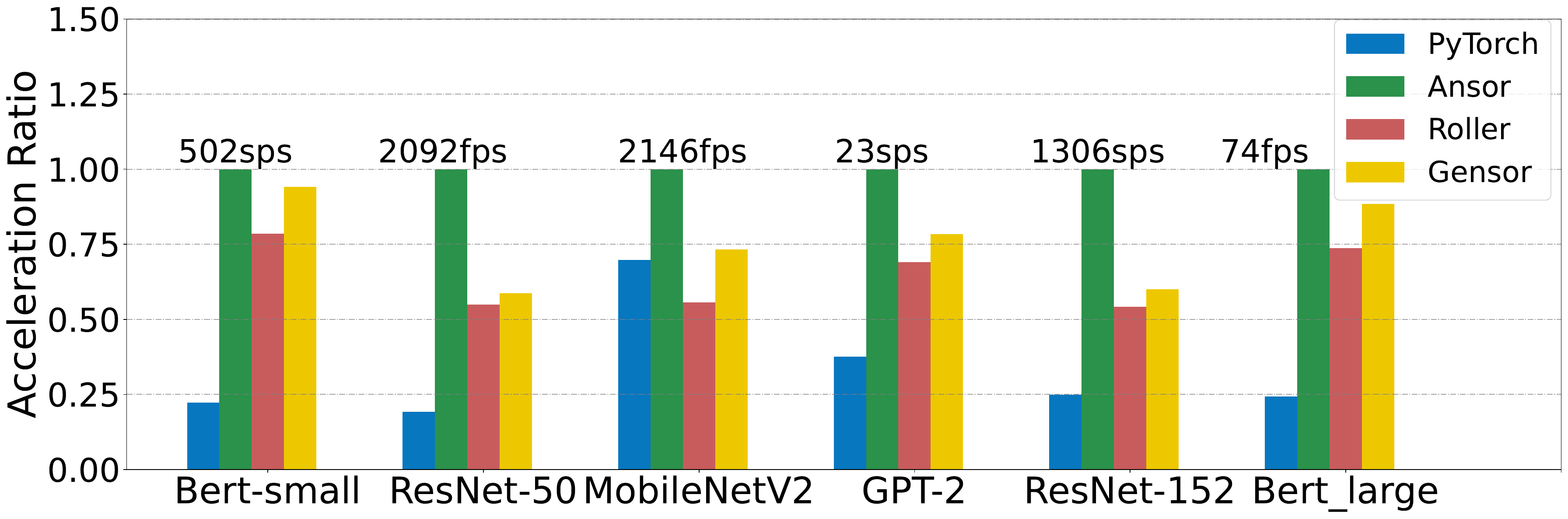}
        \caption{Performance on the RTX 4090 GPU.}
        \label{fig:ee_model_a}
    \end{subfigure}
    \vfill
    \begin{subfigure}[b]{0.85\linewidth}
        \centering
        \includegraphics[width=\linewidth]{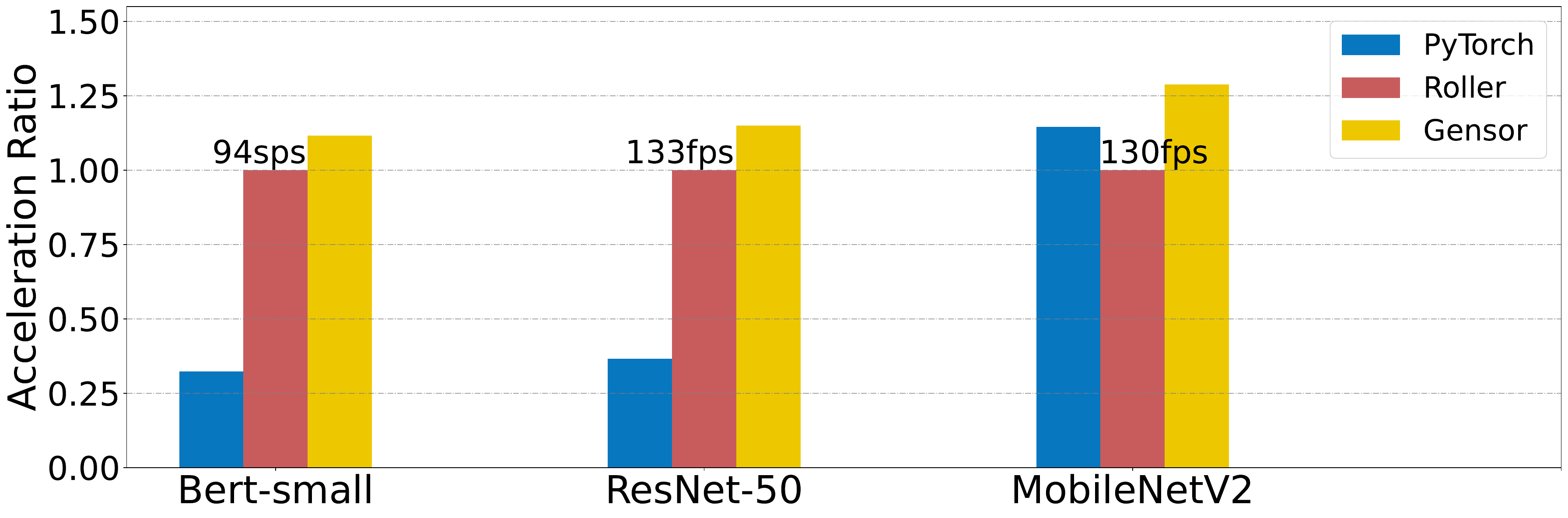}
        \caption{Performance on the Orin Nano GPU.}
        \label{fig:ee_model_b}
    \end{subfigure}
    \caption{Performance of different deep learning models (y-axis: relative performance compared to Ansor or Roller). The numbers on the baseline bars represent the absolute performance, indicated in frames or samples per second (fps/sps).}
    \label{fig:ee_model}
\end{figure}

Fig.~\ref{fig:ee_model} presents the enhancement in inference speed achieved by Gensor when applied to DNN models. We select four representative models in deep learning to evaluate acceleration performance: Bert-small\cite{kenton2019bert}, ResNet-50\cite{he2016deep}, MobileNetV2\cite{sandler2018mobilenetv2}, and GPT-2\cite{radford2019language}. 

Ansor serves as the baseline for comparing the performance of the PyTorch official implementation, Roller, and Gensor on the RTX 4090 GPU. Due to insufficient memory on edge devices, the searching method Ansor is unable to search, and the GPT-2 model is unable to run. Therefore, we choose Roller as the baseline, then run Bert-small, ResNet-50, and MobileNetV2 on the Orin Nano. The official implementations of the models are from PyTorch 2.0 repository\cite{pytorchvision}.

\begin{figure}[ht]
    \centering
    \includegraphics[width=0.8\linewidth]{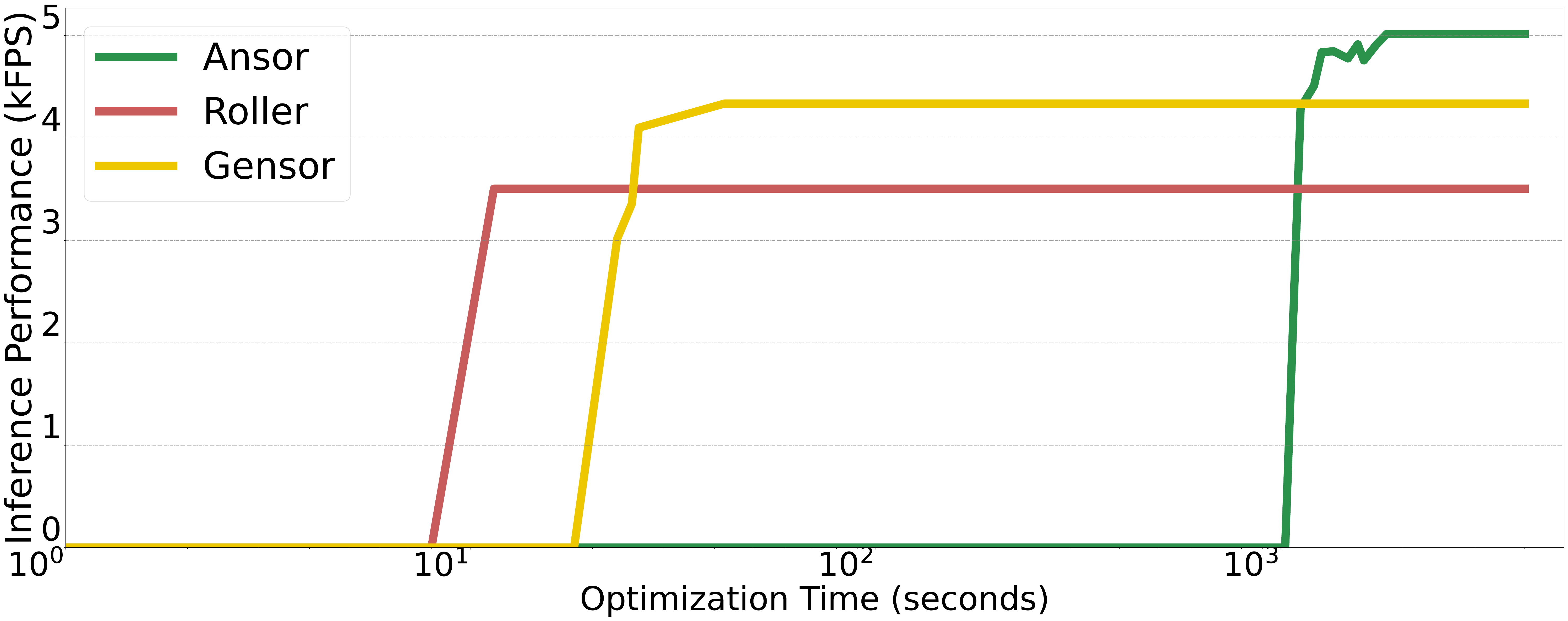}
    \caption{Comparison of different methods on the relationship between the inference performance and optimization time. The optimization time of Gensor is the same order of magnitude as Roller, yet much faster than Ansor. Furthermore, Gensor's performance is significantly better than that of Roller and is close to that of Ansor. The example is tested on ResNet-34 with an input of [128,3,224,224] on the RTX 4090 GPU.}
    \label{fig:methods_compare}
\end{figure}

As shown in Fig.~\ref{fig:ee_model}, experiments on the end-to-end models indicate that Gensor greatly improves the model inference speed, which is 1.2x the Roller speed on average, 7.2x the PyTorch implementation speed, and comparable with Ansor in Bert-small on the RTX 4090. Additionally, on the Orin Nano, Gensor is 1.19x the Roller speed on average, 2.6x the PyTorch implementation speed.

Our validation reveals that the acceleration effect of Gensor consistently outperforms Roller, the tree-based construction compilation method while being comparable to Ansor. Moreover, Gensor's significantly faster compilation time suggests it is a more efficient optimization strategy, as depicted in Fig.\ref{fig:methods_compare}. This efficiency is attributed to a superior space exploration method based on analysis rather than learning, enabling it to find optimal solutions more quickly.

\begin{figure}[ht]
    \centering
    \includegraphics[width=0.9\linewidth]{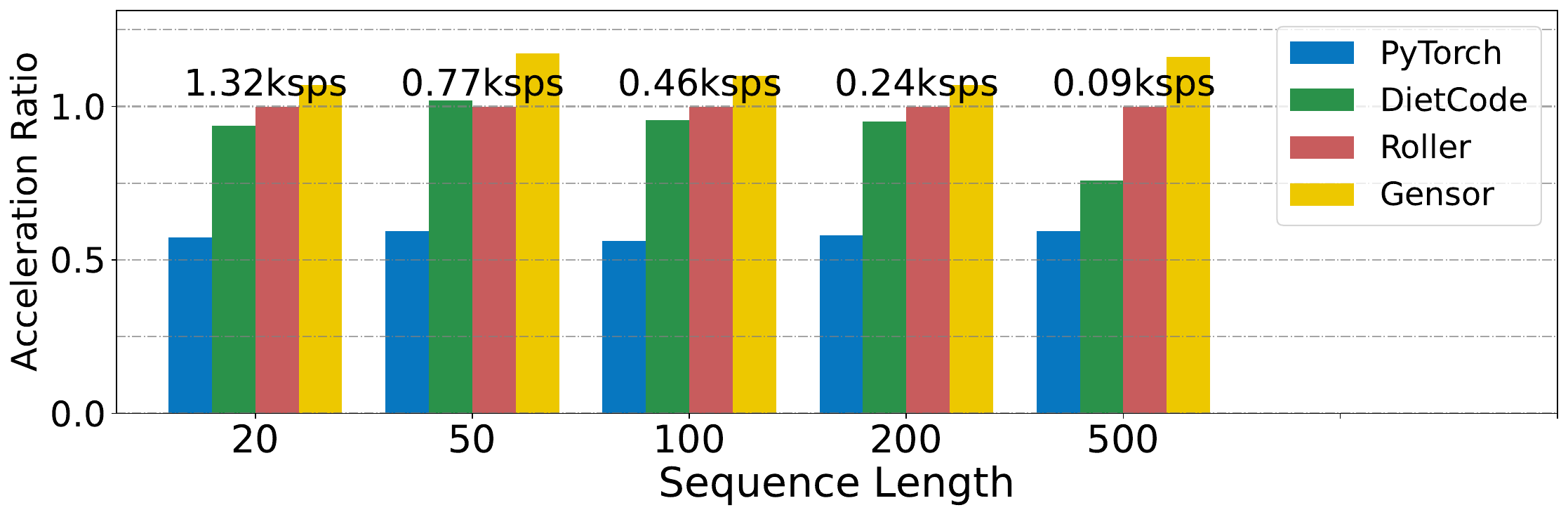}
    \caption{Bert performance with dynamic shapes (y-axis: relative performance compared to Roller). The numbers on the Roller bars represent the absolute performance, measured in kilo samples per second (ksps).}
    \label{fig:bert_dynamic}
\end{figure}

Fig.~\ref{fig:bert_dynamic} presents the inference acceleration achieved by Gensor when applied to dynamic deep learning models. We select the Bert-small with different shapes for evaluation. We choose Roller as the baseline to compare its performance with Gensor, the official implementation (PyTorch) and dynamic tensor programs optimization method, DietCode\cite{zheng2022dietcode}.
As shown in Fig.~\ref{fig:bert_dynamic}, experiments indicate that Gensor significantly improves the dynamic model inference speed, which is 1.17x faster than the Roller and 2.1x faster than the official implementation on average. The total auto-scheduling time of DietCode is about 50 minutes, and the optimization time of Gensor is about 75 minutes. Although DietCode's optimization speed is faster than Gensor's, the optimized model's performance reaches only 83\% of Gensor's. Along with the experiments in \ref{ssec:comp_time}, Gensor can achieve high-performance optimization of operators on models with dynamic input shapes in seconds.

\begin{figure*}[ht]
    \centering
    \includegraphics[width=1\linewidth]{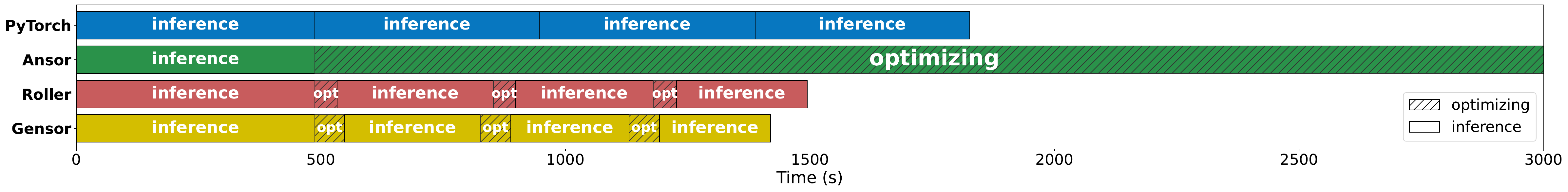}
    \caption{Optimizing and inference time of different optimization methods in dynamic structural model inference. Taking MobileNetV2 as an example, each inference stage processes 2000 times of images with a size of [128,1,224,224]. The number of channels is dynamically adjusted three times. The total time spent on optimizing and inference using Gensor is the shortest.}
    \label{fig:methods_model}
\end{figure*}

Fig.~\ref{fig:methods_model} presents the optimization and inference time of different optimization methods in dynamically adjusted model inference scenarios. We conduct simulations in a typical edge inference setting where the number of channels of the DNN is dynamically modified, with optimization carried out after each modification. In the experiment, the model first infers a fixed number of frames. Then, we modify the model structure and optimize it using different methods. This inference, modification, and optimization cycle is repeated three times. We compare the total time of directly using PyTorch without additional optimizing (resulting in zero optimization time), using the searching optimization method, Ansor, and the tree-based construction optimization method, Roller. Due to Ansor's prolonged optimization time, it is not fully displayed in the figure. Compared to other methods, Gensor exhibits the shortest total time for optimization and inference, reflecting its effectiveness in optimizing dynamic structural model inference. In practical applications, shorter compilation time can lead to faster iterations during the model development and tuning phases, which is valuable for practitioners.

Consequently, Gensor demonstrates powerful cross-platform performance in parallel processing of tensors, especially on resource-constrained edge devices. It significantly enhances the speed of operator execution and DNN inference, which is essential for executing models. Gensor also reduces compilation time, significantly speeding up the optimization cycle.


\label{ssec:ee_perf}

\section{Related Work and Discussion}
The continuous evolution of deep learning technology has led to various applications\cite{wang2023prevent}\cite{voulodimos2018deep}. Techniques such as tensor program optimization and other methods\cite{dai2023sketch} are constantly being developed to improve deep learning models.

\textbf{Manual Tensor Program Optimization: }Manual techniques have been crucial for achieving peak performance on diverse architectures for AI tasks. A notable example is the hand-tuned implementation of matrix multiplication in the Basic Linear Algebra Subprograms (BLAS)\cite{blackford2002updated}, where meticulous optimization of loop orders and tiling is conducted to align with specific processor characteristics. Similarly, manually optimized kernel libraries, like cuBLAS, show the potential of expert-crafted code to fully exploit GPU parallelism. Furthermore, the solution from tensor optimization can guide the design of underlying hardware\cite{chang2023hdsuper}.


Manual optimization, while powerful, presents significant challenges. The complexity of modern hardware and the need for deep technical expertise make it a laborious process, often requiring extensive fine-tuning effort to identify performance bottlenecks. This has led to a growing interest in automated optimization techniques, such as those employed by the TVM stack, which aims to generate high-performance code across multiple platforms without manual intervention. 

\textbf{Automatic Tensor Program Optimization (Tensor Compilation): }Most tensor compilers use optimization with searching methods\cite{dong2023akgf}. They build a huge scheduling space, then use heuristic algorithms to search for tensor programs with high performance. By learning from a corpus of optimized tensor computations, these automated systems aim to abstract the optimization process, making high-performance computing more accessible and reducing the time-to-solution for developers. TVM proposes an abstract intermediate representation method for describing tensor programs called Tensor Expression (TE)\cite{chen2018tvm}. 
AutoTVM designs a machine learning-based automatic optimization method using handwritten templates\cite{chen2018learning}. Ansor expands the search space and uses the sketch annotation method to efficiently obtain an optimized combination of scheduling primitives from a more extensive search space than autoTVM\cite{zheng2020ansor}. 
These methods make the operators generated by automatic optimization comparable to or even better than vendor libraries, greatly reducing the cost of manual optimization. However, their major problem is that the optimization time is too long due to thousands of steps during the inefficient search process.

Therefore, construction tensor compilation methods are proposed to accelerate the optimization process. Roller focuses on tensor shapes aligning with the features of the underlying processors' units. Roller then adopts a tree-based recursive algorithm to construct tensor programs\cite{zhu2022roller}. These methods construct tensor programs directly without searching, increasing the speed of tensor compilation by orders of magnitude. However, the tree structure with one single objective prevents the optimization methods from constructing high-performance tensor programs. Roller are barely comparable to vendor libraries only in a few cases.
On the contrary, Gensor significantly improves the performance of construction tensor compilation. This improvement is primarily due to the enhanced diversity in the optimization process, which results from applying a graph structure with multiple objectives.

Gensor achieves an effective balance between optimizing speed and high performance of inference. Meanwhile, in cases where the granularity of the computing platform API is too high to enable loop scheduling at the hardware level, Gensor is unsuitable. Instead, directly using micro-kernel approaches and calling vendor-provided APIs may be useful.

\section{Conclusion}

We introduce a novel graph-based construction tensor compilation method, Gensor, which outperforms existing tree-based methods. Gensor employs Markov analysis to construct tensor programs as tensor programs exhibit independent and memory-less properties. We define the state space and actions as the nodes and edges of the construction graph with the enhanced tensor IR. We design the state transition probabilities based on the tensor program's current performance and hardware architecture properties. Gensor effectively balances optimization speed and performance, generating higher-performance kernels in less time. Gensor demonstrates an exceptional ability to optimize parallel tensor programs. Extensive experiments demonstrate that Gensor outperforms the state-of-the-art construction tensor compiling method for both operators and end-to-end DNN models by 1.18x and 1.2x, respectively. Moreover, Gensor's optimization time is significantly faster than that of the state-of-the-art searching tensor compilation method. Gensor provides a more efficient and flexible approach for AI developers to accelerate their models, facilitating broad applications of AI innovations. Ongoing work aims to design a dynamic optimizing system based on Gensor to achieve efficient real-time optimization of dynamic deep neural networks. 

\section*{Acknowledgment}

This work is partially supported by the Chinese Academy of Sciences Project for Young Scientists in Basic Research (YSBR-107) and China NSF grant No.62072434.

{
\tiny
\bibliographystyle{unsrt}
\bibliography{bibliography}
}

\end{document}